\documentclass[prb,nopacs,twocolumn,preprintnumbers,amsmath,amssymb]{revtex4}

\usepackage{epsfig}
\usepackage{graphicx}
\usepackage{dcolumn}
\usepackage{color}
\usepackage{bm}
\newcommand{\be}{\begin{equation}}
\newcommand{\ee}{\end{equation}}

\begin{document}

\title{Spin-valley relaxation and quantum transport regimes \\
in two-dimensional transition metal dichalcogenides}

\author{Hector Ochoa,$^{1,2}$ Francesca Finocchiaro,$^{1}$  Francisco Guinea,$^{1,3}$ Vladimir I. Fal'ko$^{4}$}

{\affiliation{$^1$ Instituto de Ciencia de Materiales de Madrid, CSIC, 28049 Madrid, Spain
\\
$^2$ Donostia International Physics Center (DIPC),
20018 San Sebasti\'an, Spain
\\
$^3$ School of Physics ad Astronomy, University of Manchester, Oxford Road, Manchester, M13 9PL, UK
\\
$^4$Physics Department, Lancaster University, Lancaster, LA1 4YB, UK}

\begin{abstract}
Quantum transport and spintronics regimes are studied in p- and n-doped atomic layers of hexagonal transition metal dichalcogenides (TMDCs), subject to the interplay between the valley structure and spin-orbit coupling. We find how spin relaxation of carriers depends on their areal density and show that it vanishes for holes near the band edge, leading to the density-independent spin diffusion length, and we develop a theory of weak localisation/antilocalisation, describing the crossovers between the orthogonal, double-unitary and symplectic regimes of quantum transport in TMDCs.
\end{abstract}

\maketitle

\section{Introduction}

Among the novel two-dimensional materials that have attracted a lot of attention during the recent years, semiconducting transition metal dichalcogenides (TMDCs)\cite{TMDC} are particularly interesting due to their potential for applications in electronics and optoelectronics.\cite{opt1,opt2,opt3,opt4,opt5} They have been recently implemented in field-effect transitors,\cite{TMDC1,TMDC2,TMDC3,Fang_etal,Iate_etal,Liu_etal,Kang_etal,Das_etal} showing large in-plane mobilities and a high current on/off ratios, which make them also very interesting for sensors. This has motivated a big effort in the study of the transport properties of these single-layer crystals.\cite{mob1,mob2,mob3,mob4,Feng_etal,Ovchinnikov_etal}

Contrary to their bulk counterparts, single-layer TMDCs are considered to have a direct band-gap,\cite{Roldan_etal_rev} which appears at the corners $K_{\pm}$ of the hexagonal Brillouin zone (see Fig.~\ref{fig:lattice}). Simultaneously, the large spin-orbit (SO) interaction provided by the heavy transition metal atoms together with the lack of inversion symmetry splits the energy bands around $K_{\pm}$ points,\cite{two_band,2DBands3,2DBands4,2DBands5,arpes1,arpes2,spinorbit1,spinorbit2} where the out-of-plane spin polarization is still a good quantum number due to the mirror symmetry ($z\rightarrow-z$) of the system. The interplay between valley and spin degrees of freedom influences charge and spin transport characteristics of these materials. In general, magnetotransport experiments in systems with large SO coupling provide insights about the nature of momentum scattering and spin relaxation processes, whereas the interplay between spin-lattice relaxation and quantum transport leads to a crossover between orthogonal and symplectic classes of quantum disordered systems,\cite{RMT1,RMT2} manifested in measurements as weak localisation\cite{WL} (WL) and weak anti-localisation\cite{WAL} (WAL) magnetoresistance (MR). This interplay acquires an additional twist in 2D conductors with a multi-valley band structure.\cite{IK,Guinea,MLG,BLG,lu} In the case of TMDCs, the SO spin splitting tends to suppress spin relaxation, whereas lattice defects and deformations mimic time-inversion symmetry breaking for the intravalley propagation of carriers (though the true time-inversion symmetry is preserved because it involves interchanging the valleys).

\begin{figure}
\includegraphics[width=1\columnwidth]{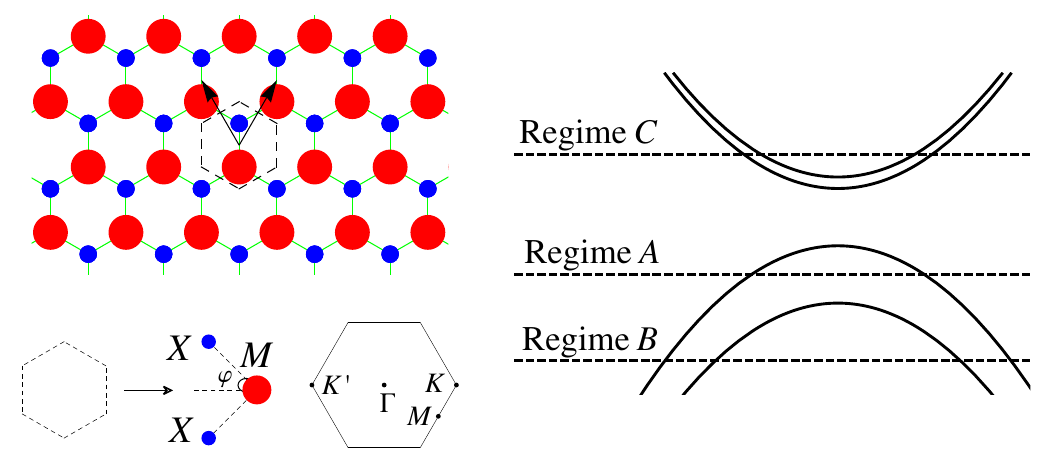}
 \caption{Left: Lattice, unit cell and Brillouin zone of TMDC monolayers. Right: Schematic view of the quantum transport regimes discussed through the text.}
\label{fig:lattice}
\end{figure}

In this work, we study in detail the interplay between SO coupling and multi-valley properties in spintronics and quantum transport of single-layer TMDCs. We identify three regimes of quantum transport, which are distinguished by the relative size of the SO splitting and the Fermi energy of charge carriers, as scketched in Fig.~\ref{fig:lattice}. The structure of the manuscript is the following: In Sec.~II we present the phenomenological model for TMDCs band structure and disorder, supported by $\mathbf{k}\cdot\mathbf{p}$ theory derivation in Appendices~\ref{sec:AppA} and \ref{sec:AppB}. In Sec.~III, we analyze the interference correction to the conductivity and show that it displays various forms of the WL to WAL crossover, depending on the relation between the spin splitting and Fermi energy of the carriers. In Sec.~IV, we discuss the ocurrence of such crossovers in e.g. MoS$_2$. 

\section{The model}

Two-dimensional unit cells of TMDCs consist on $X$-$M$-$X$ layers, where the transition metal atoms ($M$) are ordered in a triangular lattice, each of them bonded to six chalcogen atoms ($X$) located in the top and bottom layers, see Fig.~\ref{fig:lattice}. We focus on the dynamics of spinful electrons and holes around the $K_{\pm}$ valleys. To that end, we introduce Pauli algebras associated to valley ($\tau$ matrices) and spin ($s$ matrices), which can be classified according to the irreducible representations of $D_{3h}''$, see Tab.~\ref{tab:electronic_operators_main}. Here we deal with $D_{3h}''$, the point group associated to the tripled unit cell, because it allows us to treat excitations at both valleys on an equal footing, see Appendix~\ref{sec:AppA}.

The $\mathbf{k}\cdot\mathbf{p}$ theory Hamiltonian for electrons and holes in TMDCs has the form\cite{2DBands1,2DBands2,2DBands3,2DBands4,2DBands5,Kormanyos_etal_rev,footnotemass}
\begin{subequations}
\label{Hamiltonian}
\begin{align}
\mathcal{H}=\frac{\left|\mathbf{p}\right|^2}{2m^*}+
\mu\left(p_x^3-3p_xp_y^2\right)\tau_z+\frac{\lambda}{2} \tau_z s_z
+\delta\mathcal{H}\left(\mathbf{r}\right).
\label{Ham}
\end{align}
This takes into account trigonal warping, $\mu$, of their dispersion inverted in $K_{\pm}$ valleys and SO splitting, $\lambda$ (large/small in the valence/conduction band, see Tab.~\ref{tab:materials}). The last term in Eq.~(\ref{Ham}),
\begin{widetext}
\begin{gather}
\delta\mathcal{H}\left(\mathbf{r}\right)=
u_0\left(\mathbf{r}\right)+
u_{z}\left(\mathbf{r}\right)\tau_{z} s_{z}+
\left\{\mathbf{p},\mathbf{a}_g\left(\mathbf{r}\right)\right\}\tau_z
+\left\{\mathbf{p},\mathbf{a}_{gz}\left(\mathbf{r}\right)\right\}s_{z}
+\mathbf{u}_{sf}\left(\mathbf{r}\right)\cdot\mathbf{s}\tau_{z}+
\sum_{\alpha=x,y}\left\{\mathbf{p},\mathbf{w}_{\alpha}\left(\mathbf{r}\right)\right\}s_{\alpha}
\nonumber\\
+\mathbf{u}_i\left(\mathbf{r}\right)\cdot
\boldsymbol{\tau}+
\sum_{\alpha=x,y}\left\{\mathbf{p},\mathbf{w}_{z\alpha}\right\}\tau_{\alpha}s_z+
\sum_{\alpha,\beta=x,y}\left\{\mathbf{p},\mathbf{w}_{\alpha\beta}\right\}\tau_{\alpha}s_{\beta},
\label{eq:dis}
\end{gather}
\end{widetext}
\end{subequations}
describes imperfections in the 2D crystal that, in principle, break all its symmetries except for time-reversal.

The first two terms in $\delta\mathcal{H}\left(\mathbf{r}\right)$ stand for intravalley disorder, sensitive to the allowed spin state of the electron in each valley. The next two terms account for both lattice deformations (responsible for a valley/spin-dependent pseudo-magnetic field\cite{IK,Guinea}) and the Berry curvature specific for the bands at the corners of the Brillouin zone. Their $\mathbf{k}\cdot\mathbf{p}$ theory\cite{2DBands5} derivation is described in Appendix~\ref{sec:AppB}. The presence of the last two terms in the first line, with spin operators $\left(s_x,s_y\right) \equiv \mathbf{s}$, requires $z\rightarrow-z$ symmetry breaking, e.g.,  by flexural deformations of the 2D crystal in the case of the last term.\cite{ochoa_etal} 

The second line in Eq.~\eqref{eq:dis}, with valley Pauli matrices $\left(\tau_x,\tau_y\right) \equiv \boldsymbol{\tau}$, describes intervalley disorder due to atomic defects in the crystal. The first two terms account for intervalley scattering without spin-flip; the last term represents the only intervalley spin-flip perturbation permited by the time-inversion symmetry in the lowest-order $\mathbf{k}\cdot\mathbf{p}$ expansion around $K_{\pm}$. The momentum dependence of such a term suggests that the intervalley spin-flip scattering is absent for the carriers at the band edge.\cite{foot,Yafet}

\begin{center}
\begin{table}
\begin{tabular}{|c||c|c|}
\hline
Irrep&$t\rightarrow -t$ invariant&$t\rightarrow -t$ odd\\
\hline
\hline
$A_2'$&&$\tau_z$, $s_z$\\
\hline
$E''$&&$\left(\begin{array}{c}
s_x\\
s_y
\end{array}\right)$\\
\hline
$E_{1,2}'$&$\left(\begin{array}{c}
\tau_x\\
\tau_y
\end{array}\right)$&\\
\hline
\end{tabular}
\caption{Definitions of valley and spin matrices.}
\label{tab:electronic_operators_main}
\end{table}
\end{center}

\section{Quantum transport regimes}

\begin{center}
\begin{table}
\begin{tabular}{|c|c|c|c|c|}
\hline
Material&Band&$\frac{m^{*}}{m_0}$&$\lambda$ (meV)&Mobility (cm$^2$/sV)\\
\hline
\hline
&e$^-$&0.46 [\onlinecite{2DBands5,Kormanyos_etal_rev}]&3 [\onlinecite{2DBands5,Kormanyos_etal_rev}] &20-350
 [\onlinecite{TMDC1,Iate_etal,mob1,mob2,mob3,mob4}]\\
\cline{2-5}
MoS$_2$&h$^+$&0.54 [\onlinecite{2DBands5,Kormanyos_etal_rev}]&148 [\onlinecite{2DBands5,Kormanyos_etal_rev}]&\\
\hline
&e$^-$&0.56 [\onlinecite{Kormanyos_etal_rev}]&22 [\onlinecite{Kormanyos_etal_rev}]&\\
\cline{2-5}
MoSe$_2$&h$^+$&-0.59 [\onlinecite{Kormanyos_etal_rev}]&186 [\onlinecite{Kormanyos_etal_rev}]&\\
\hline
&e$^-$&0.26 [\onlinecite{Kormanyos_etal_rev}]&-32 [\onlinecite{Kormanyos_etal_rev}]&\\
\cline{2-5}
WS$_2$&h$^+$&-0.35 [\onlinecite{Kormanyos_etal_rev}]&430 [\onlinecite{two_band,Feng_etal,Kormanyos_etal_rev}]&50-120 [\onlinecite{Ovchinnikov_etal}]\\
\hline
&e$^-$&0.28 [\onlinecite{Kormanyos_etal_rev}]&-37 [\onlinecite{Kormanyos_etal_rev}]&\\
\cline{2-5}
WSe$_2$&h$^+$&-0.36 [\onlinecite{Kormanyos_etal_rev}]&460 [\onlinecite{two_band,Feng_etal,Kormanyos_etal_rev}]&140 [\onlinecite{Fang_etal}]\\
\hline
\end{tabular}
\caption{Effective mass, SO splitting and mobilities for conduction (e$^-$) and valence (h$^{+}$) bands of some best-studies semiconducting TMDCs.}
\label{tab:materials}
\end{table}
\end{center}

We study the phase-coherent quantum interference correction to conductivity since the typical mean-free paths deduced from the mobilities reported in monolayers of TMDCs (see Tab.~\ref{tab:materials}) are of the order of $\ell\sim1-10$ nm, whereas the phase-coherence lengths reported in magnetotransport experiments in few-layer samples are of the order of $\ell_{\varphi}\sim50-100$ nm.\cite{MR1,MR2} The analysis of quantum transport characteristics of
TMDCs is  based on the diagramatic perturbation theory calculations similar to those performed earlier in graphene\cite{MLG,BLG,KKF} and other 2D materials.\cite{lu} Depending on the relative size of SO splitting and Fermi energy $\varepsilon_F$ of charge carriers, see Fig.~\ref{fig:lattice}, we identify three distinct spin/valley relaxation and quantum transport regimes:  
\begin{itemize}
\item[\textbf{A.}] $\boldsymbol{\lambda>\varepsilon_F}$: lightly p-doped monolayers of MoX$_2$ and WX$_2$ (X=S, Se, Te) with holes fully spin-polarized in opposite directions in the opposite valleys.\cite{arpes1,arpes2,spinorbit1,spinorbit2}
\item[\textbf{B.}] $\boldsymbol{\varepsilon_F\gtrsim\lambda}$: specific for heavily p-doped MoS$_2$.
\item[\textbf{C.}] $\boldsymbol{\varepsilon_F \gg \lambda}$: typical for n-doped MoX$_2$ monolayers.
\end{itemize}

\begin{figure}
\includegraphics[width=0.4\columnwidth]{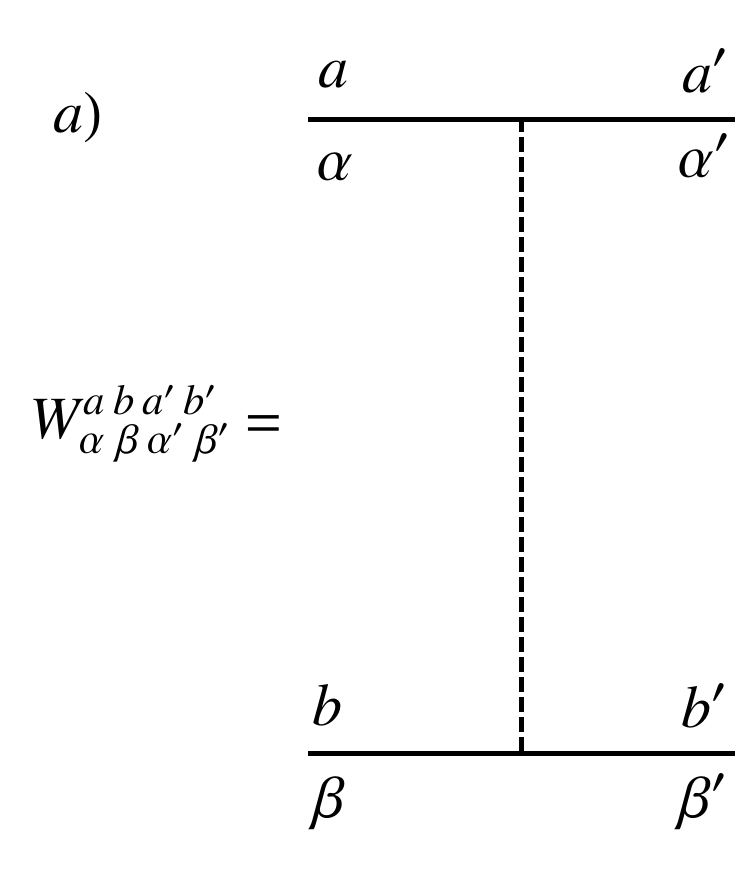}
\includegraphics[width=0.55\columnwidth]{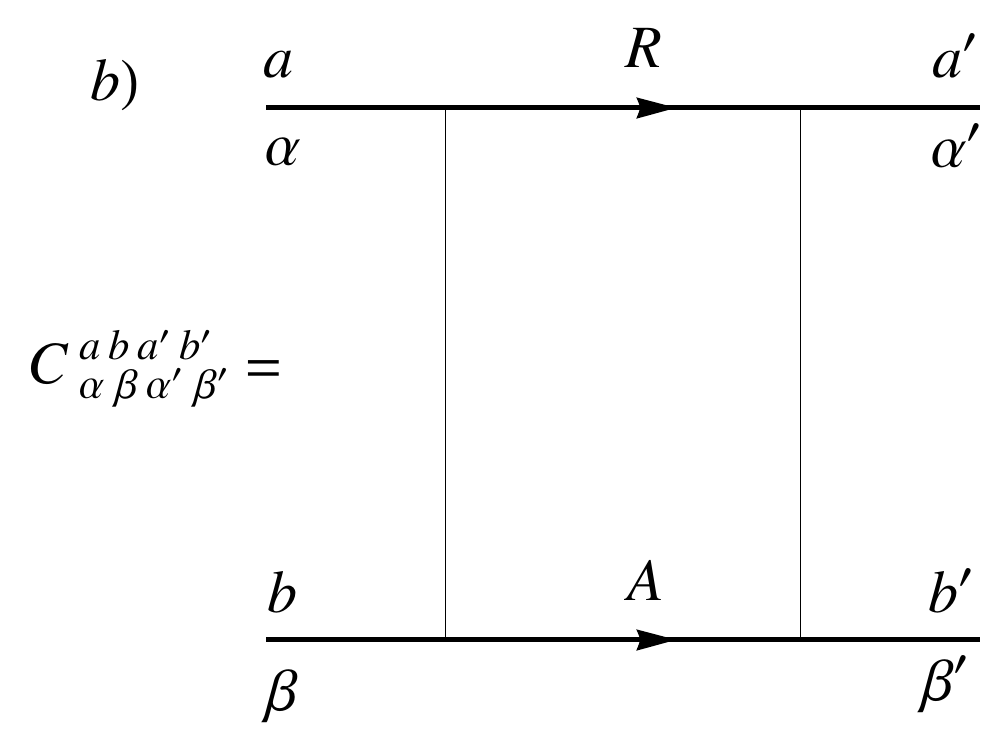}\\
\includegraphics[width=1\columnwidth]{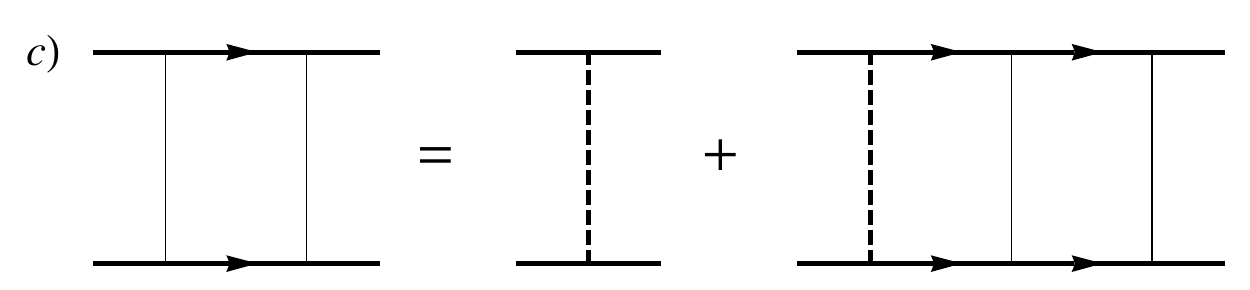}
 \caption{Diagramatic representation of a) disorder correlator, b) Cooperon, and c) Bethe-Salpeter equation.}
\label{fig:diagrams}
\end{figure}

\subsection*{ Regime A: $\boldsymbol{\lambda>\epsilon_F}$ (valence band)}

Energy conservation and spin polarization of electrons in opposite directions ($\uparrow\downarrow$) in valleys $K_{\pm}$ do not leave any space for intravalley spin-flip and intervalley spin-conserving scatterings. This makes redundant the last two terms in the first line of Eq.~\eqref{eq:dis} and first two terms in the second line. Then, spin-conserving intravalley disorder is characterized by the scattering rate,
\begin{subequations}
\begin{gather}
\tau_{0}^{-1} = \frac{2\pi\nu(\Omega_{0}+\Omega_{z})}{\hbar},
\qquad \nu=\frac{m^*}{2\pi\hbar^2},
\label{0s}
\\
\left\langle u_{\alpha}\left(\mathbf{r}\right)u_{\beta}\left(\mathbf{r}'\right)\right\rangle=
\Omega_{\alpha}\delta_{\alpha\beta}\delta\left(\mathbf{r}-\mathbf{r}'\right), \quad \alpha=(0,z).
\nonumber
\end{gather}
The gauge-field-like part of $\delta\mathcal{H}$ determines the rate, 
\label{linear}
\begin{gather}
\tau_{g}^{-1}=\frac{2\pi\nu p_F^2(\Theta_{g}+\Theta_{gz})}{\hbar}\propto n_h, 
\label{gaugescattering}
\\
\left\langle a_{\alpha}^{i}\left(\mathbf{r}\right)a_{\beta}^{j}\left(\mathbf{r}'\right)\right\rangle=
\Theta_{\alpha}\delta_{\alpha\beta}\delta_{ij}\delta\left(\mathbf{r}-\mathbf{r}'\right), \quad \alpha=(g,gz),
\nonumber 
\end{gather}
which scale linearly with the hole density, $n_h$. 
The last term in Eq.~\eqref{eq:dis}, with $\left(\alpha,\beta=x,y\right)$ is responsible for the only possible inter-valley scattering process in the regime A, accompanied by a spin-flip, which determines the hole spin relaxation rate,\cite{foot}
\begin{gather}
\tau_{is}^{-1}=\frac{8\pi\nu p_F^2\Theta_{is}}{\hbar} \propto n_h, 
\label{holespinrelax}
\\
\left\langle w_{\alpha\beta}^i\left(\mathbf{r}\right)w_{\alpha'\beta'}^j\left(\mathbf{r}'\right)\right\rangle=
\Theta_{is}\delta_{\alpha\alpha'}\delta_{\beta\beta'}\delta_{ij}
\delta\left(\mathbf{r}-\mathbf{r}'\right),
\nonumber
\end{gather}
\end{subequations}
which also scales linearly with the hole density. 

Rates (\ref{linear}) sum up into the momentum relaxation rate, 
$$\tau^{-1}=\tau_{0}^{-1}+\tau_{g}^{-1}+\tau_{is}^{-1},$$
which determines the value of Drude conductivity and diffusion coefficient, $D=\frac12 \tau v_F^2$, and the result of Eq.~(\ref{holespinrelax}) suggests that in p-doped TMDCs spin-diffusion lengths 
\begin{align}
L_{is}^{(a)}=\sqrt{D\tau_{is}} \sim \sqrt{\hbar^3\tau/2\Theta_{is} m_*^3},
\label{Ls}
\end{align}
are almost independent of the carrier density at $n_h\rightarrow0$.

The corrections due to quantum interference are exppresed in terms of particle-particle correlation functions known as Cooperons, whose dynamics are governed by the Bethe-Salpeter equation, diagramatically depicted in Fig.~\ref{fig:diagrams}. Since spin and valley degrees of freedom are coupled in this regime, we simplify the notation by introducing a set of generators of U(2), $\sigma_{0,x,y,z}$, with $\sigma_0$ the identity and $\sigma_{x,y,z}$ Pauli matrices acting on the Hilbert space span by the doublet ($\mathbf{K}_{+},\uparrow$), ($\mathbf{K}_{-},\downarrow$).\cite{foot2} First, we decompose the disorder correlators introduced in the main text (Fig.~\ref{fig:diagrams}~a)) and Cooperons (Fig.~\ref{fig:diagrams}~b)) in singlet ($l,s=0$) and triplet ($l,s=x,y,z$) modes as
\begin{gather*}
C_{ss'}\equiv\frac{1}{2}\left[\sigma_y\sigma_s\right]_{\alpha\beta}
C_{\alpha\beta\alpha'\beta'}
\left[\sigma_{s'}\sigma_y\right]_{\beta'\alpha'},
\nonumber\\
W_{ss'}\equiv\frac{1}{2}\left[\sigma_y\sigma_s\right]_{\alpha\beta}
W_{\alpha\beta\alpha'\beta'}
\left[\sigma_{s'}\sigma_y\right]_{\beta'\alpha'},
\end{gather*}
where the sum in $\alpha,\beta,\alpha',\beta'$ indices is assumed.
Then, the Bethe-Salpeter equations (Fig.~\ref{fig:diagrams}~c)) can be written in a
compact way as\begin{align*}
C_{s_1s_2}\left(\mathbf{Q},\omega\right)=W_{s_1s_2}+\sum_{s,s'}\sum_{l,l'} W_{s_1s'}C_{ss_2}\left(\mathbf{Q},\omega\right)\Pi_{ss'}\left(\mathbf{Q},\omega\right),
\label{eq:BS}
\end{align*}
where
\begin{widetext}
\begin{align}
\Pi_{ss'}\left(\mathbf{Q},\omega\right)\equiv\frac{1}{2}\int \frac{d^2\mathbf{p}}{\left(2\pi\hbar\right)^2} \mbox{Tr}\left[\sigma_s\sigma_y\left(\hat{G}^R\left(\mathbf{p},\hbar\omega+\varepsilon_F\right)\right)^{T}\sigma_y\sigma_{s'}
\hat{G}^A\left(\mathbf{Q}-\mathbf{p},\varepsilon_F\right)\right].
\end{align}
\end{widetext}
The retarded/avanced Green operators are just
\begin{align}
\hat{G}^{R,A}\left(\mathbf{p},\omega\right)=\frac{1}{\hbar\omega-\varepsilon_{\mathbf{p}}\pm i\frac{\hbar}{2\tau}}\sigma_0,
\end{align}
so then we have\begin{align}
\Pi_{ss'}\left(\mathbf{Q},\omega\right)=\approx\frac{2\pi\nu\tau}{\hbar}\left(1+i\tau\omega-\tau D\left|\mathbf{Q}\right|^2\right)\times\delta_{ss'},
\label{eq:po1}
\end{align}
where the last result corresponds to the polarization operator in the so-called {\it diffusive approximation} ($\tau D\left|\mathbf{Q}\right|^2,\tau\omega\ll 1$).

The WL correction to conductivity can be written as the contribution of 4 Cooperon modes (1 singlet $s=0$ and 3 triplet $s=x,y,z$, with $\tilde{C}_s\equiv\frac{2\pi\nu\tau^2}{\hbar} C_{ss}\left(\mathbf{Q}\right)$),
\begin{align}
\delta g_{ii}=-\frac{2e^2D}{\pi\hbar}\int
\frac{d^2\mathbf{Q}}{\left(2\pi\right)^2}\left[\tilde{C}_x+\tilde{C}_y+
\tilde{C}_z-\tilde{C}_0\right],
\label{eq:aux}
\end{align}
which are fundamental solutions of diffusion-relaxation kernels,\begin{align}
\left[-D\nabla^2-i\omega+\Gamma_s\right]\tilde{C}_s\left(\mathbf{r}-\mathbf{r}',\omega\right)=\delta\left(\mathbf{r}-\mathbf{r}'\right).
\end{align}
This is deduced from the Bethe-Salpeter equation, Eq.~\eqref{eq:BS}, assuming the low frequency and momentum expansion for the polarization operator, Eq.~\eqref{eq:po1}, and also that diagonal scattering dominate over the rest, $\tau/\tau_0\sim 1$.\cite{foot3}

\begin{center}
\begin{table}
\begin{tabular}{|c||c|}
\hline
Relaxation gaps&Relaxation rates\\
\hline
\hline
$\Gamma_0=0$&$\tau^{-1}=\tau_0^{-1}+\tau_{g}^{-1}+
+\tau_{is}^{-1}$\\
$\Gamma_x=\Gamma_y=\tau_{*}^{-1}+\tau_{is}^{-1}$&\\
$\Gamma_z=2\tau_{is}^{-1}$&$\tau_{*}^{-1}=2\tau_g^{-1}+\frac{15\mu^2p_F^6\tau}{4\hbar^2}$\\
\hline
\end{tabular}
\caption{Relation between Cooperon relaxation gaps and scattering rates in the regime A.}
\label{tab:gapsa}
\end{table}
\end{center}

The relations between the relaxation gaps, $\Gamma_s$, and the rates associated to different scattering mechanisms are summarized in Tab.~\ref{tab:gapsa}. Similarly to other multi-valley conductors without intervalley scattering, lattice defects\cite{IK,Guinea,MLG,BLG,KKF} and trigonal warping in the valley dispersion\cite{MLG,BLG} 
suppress the low-temperature part of the quantum correction to conductivity caused by the interference of phase-coherent diffusive waves encircling the same random walk trajectory in the reversed directions. This is because inhomogeneous deformations generate a random pseudo-magnetic field with the opposite sign in $K_{\pm}$ valleys, whereas trigonal anisotropy splits hole's wavenumber for the opposite Fermi velocity directions, hence inducing a random phase difference for the clockwise and anticlockwise propagating waves, Fig.~\ref{fig:mag_1} (with the opposite sign in the opposite valleys). The cumulative effect of these two factors determines the decay rate $\tau_*^{-1}$ of valley-polarised Cooperons\cite{MLG,BLG,KKF} in the set of triplet and singlet two-hole correlation functions, 
\begin{align}
\tau_{*}^{-1}=2\tau_g^{-1}+\frac{15\mu^2p_F^6\tau}{4\hbar^2} \propto An_h+Cn_h^3.
\label{tau*}
\end{align}
Note that, similarly to $L_{is}$ in Eq.~\eqref{Ls},
\begin{align} 
L_{*}\equiv\sqrt{D\tau_*}\sim\sqrt{\hbar^3\tau/\left(\Theta_{g}+\Theta_{gz}\right)m_*^3}
\label{L*}
\end{align}
is independent of hole densities at $n_h\rightarrow0$.

\begin{figure}
\includegraphics[width=0.49\columnwidth]{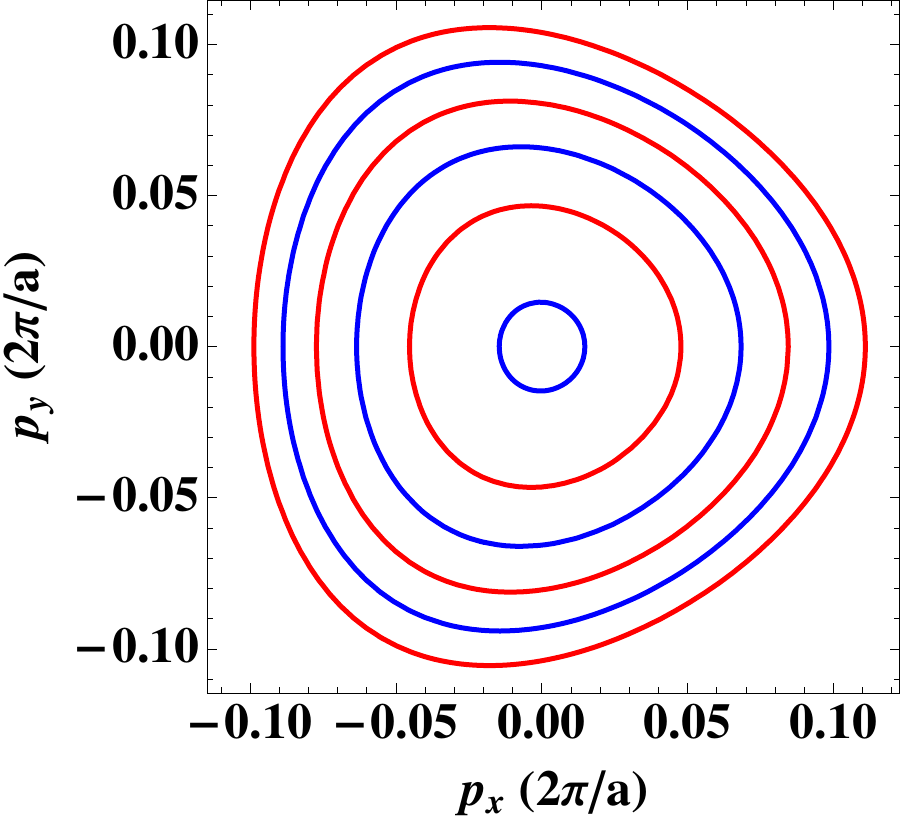}
\includegraphics[width=0.49\columnwidth]{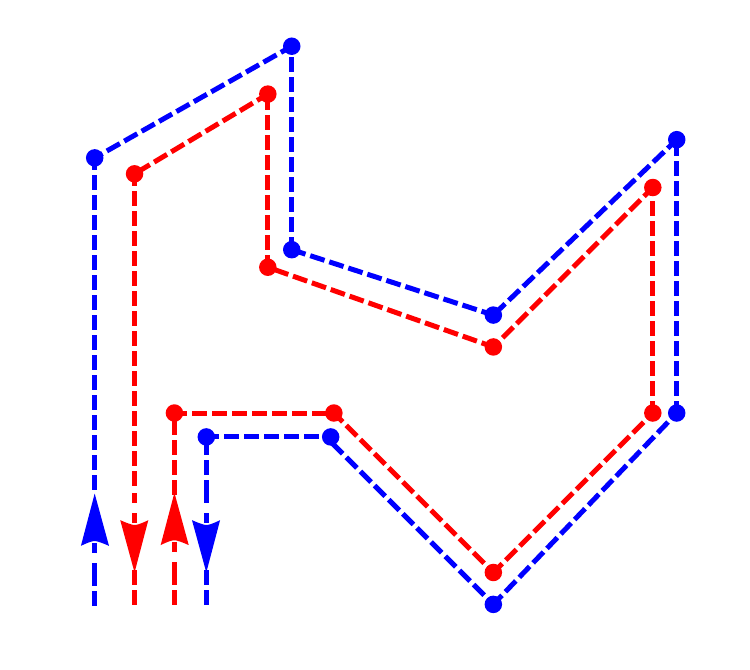}
 \caption{Trigonal warping of isoenergy contours in the valence band in one of the valleys (left), leading to a phase difference between holes propagating along a closed path in the reverse directions (right).}
\label{fig:mag_1}
\end{figure}

In the diffusive approximation, the integration in Eq.~\eqref{eq:aux} has a natural ultraviolet cut-off imposed by the inverse of the mean free path, $\ell=\sqrt{D\tau}$, whereas the infrared cut-off is imposed by the decoherence length $\ell_{\varphi}=\sqrt{D\tau_{\varphi}}$. Then, the WL correction to conductivity reads just as
\begin{gather}
\delta g=\frac{e^2}{\pi h}\left[\ln\left(\frac{\tau^{-1}}{\tau_{\varphi}^{-1}}\right)
\ln\left(\frac{\tau^{-1}}{\tau_{\varphi}^{-1}+2\tau_{is}^{-1}}\right)-
\right.
\nonumber
\\
\qquad \left. -2\ln\left(\frac{\tau^{-1}}{\tau_{\varphi}^{-1}+\tau_{is}^{-1}+\tau_{*}^{-1}}\right)\right].
\label{crossover-g}
\end{gather}
In the presence of an out-of-plane magnetic field, $B$, $\left|\mathbf{Q}\right|^2$ is quantized into $Q_n^2=\left(n+1/2\right)\ell_{B_z}^{-2}$, with the magnetic length defined as $\ell_B=\sqrt{\hbar /4eB}$. For $\ell_B\gg \ell$ the diffusive approximation is still valid, and the WL correction to conductivity reads\begin{align*}
\delta g_{ii}\left(B\right)=-\frac{e^2D\ell_B^{-2}}{\pi h}\sum_{s}\sum_{n=0}^{n_{max}}\frac{c_s}{D\ell_B^{-2}\left(n+\frac{1}{2}\right)+\Gamma_s+\tau_{\varphi}^{-1}},
\end{align*}
with $c_{0,x,y,z}=-1,+1,+1,+1$.
Thanks to the property of the digamma function,\begin{align*}
\psi\left(x+n_{max}+1\right)-\psi\left(x\right)=\sum_{n=0}^{n_{max}}\frac{1}{x+n},
\end{align*}
we can perform the summation, leading to\begin{align*}
\delta g_{ii}\left(B\right)=-\frac{e^2}{\pi h}\sum_{s}c_{s}\left[\ln\left(\frac{\hbar\tau^{-1}}
{4eDB}\right)-\psi\left(\frac{1}{2}+\frac{B_s+B_{\varphi}}{B}\right)\right],
\end{align*}
with $B_{s}=\frac{\hbar\Gamma_s}{4eD}$. Note that we have taken the limit $n_{max}\rightarrow \infty$,\begin{align*}
\psi\left(x+n_{max}+\frac{3}{2}\right)\rightarrow\ln\left(n_{max}\right)\approx\ln\left(\frac{\hbar\tau^{-1}}
{4eDB}\right).
\end{align*}
Substrating the zero field correction, the MR defined as $\Delta\rho\left(B\right)=\rho\left(B\right)-\rho\left(0\right)=-\Delta g\left(B\right)/\rho^2$ can be written as\begin{gather}
\frac{\Delta\rho\left(B\right)}{\rho^2}=\frac{e^2}{\pi h}\left[F\left(\frac{B}{B_{\varphi}}\right)
-F\left(\frac{B}{B_{\varphi}+2B_{is}}\right)
\right.
\nonumber
\\
\qquad \qquad \qquad \left. -2F\left(\frac{B}{B_{\varphi}+B_{is}+B_{*}}\right)\right],
\label{crossover-MR}
\\
F\left(z\right)\equiv\ln\left(z\right)+\psi\left(\frac{1}{2}+\frac{1}{z}\right), \quad
B_{\alpha}\equiv\frac{\hbar/e}{4D\tau_{\alpha}}.
\nonumber
\end{gather}

Without intervalley scattering, the combined effect of pseudo-magnetic disorder and trigonal warping places the system in the double-unitary symmetry class\cite{RMT2,aleiner_falko} with a WL peak saturated at a temperature such that $\tau_{\varphi}(T)\sim\tau_{*}$, and this intreplays with spin-flip intervalley scattering which drives the system to the WAL regime.  
The resulting behaviour of MR is illustrated in Fig.~\ref{fig:mag_2}. From the density dependence of relaxation rates $\tau_{is}^{-1}$ and $\tau_{*}^{-1}$, $B_{is}\propto L_{is}^{-2}$ and $B_{*}\propto L_{*}^{-2}$ should be finite at $n_h\rightarrow 0$, see Eqs.~\eqref{Ls} and \eqref{L*}, in contrast to $B_{\varphi}\propto \/n_h^{-1}$. Hence, we conclude that magnetoresistance displays a crossover from WL to WAL behavior upon increasing the hole density. Here, the form of MR would be dependent on the amount of atomic defects responsible for the spin-flip intervalley scattering: for a virtually defectless crystal, MR would display a two-step crossover, from suppressed WL to WAL (this behaviour is exactly the reverse of the WAL-WL crossover in monolayer graphene\cite{MLG}).

\begin{figure}
\includegraphics[width=0.51\columnwidth]{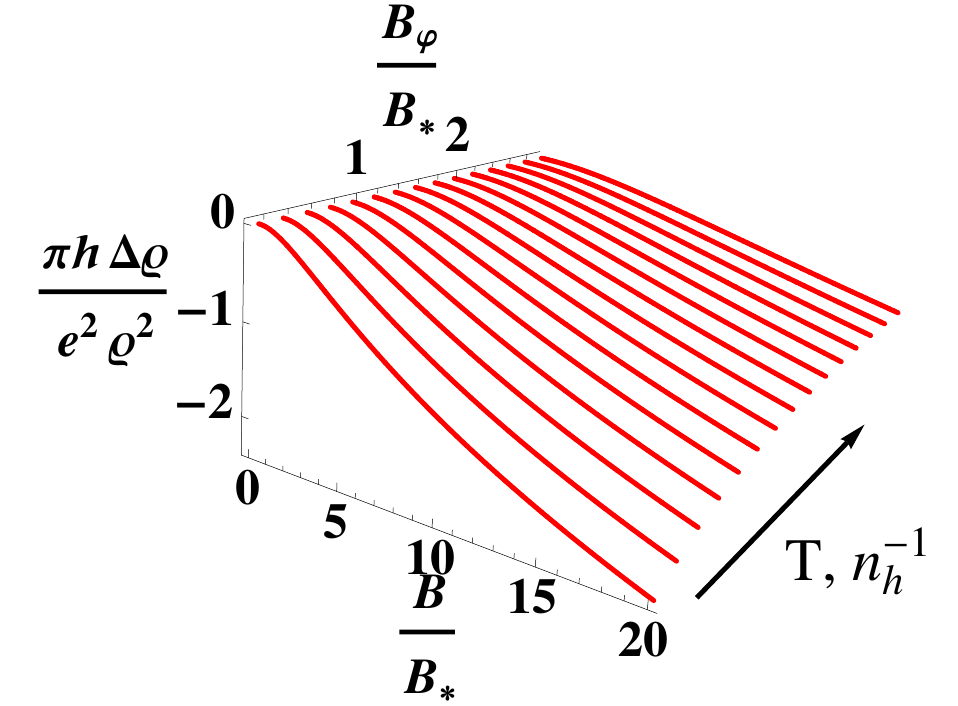}
\includegraphics[width=0.47\columnwidth]{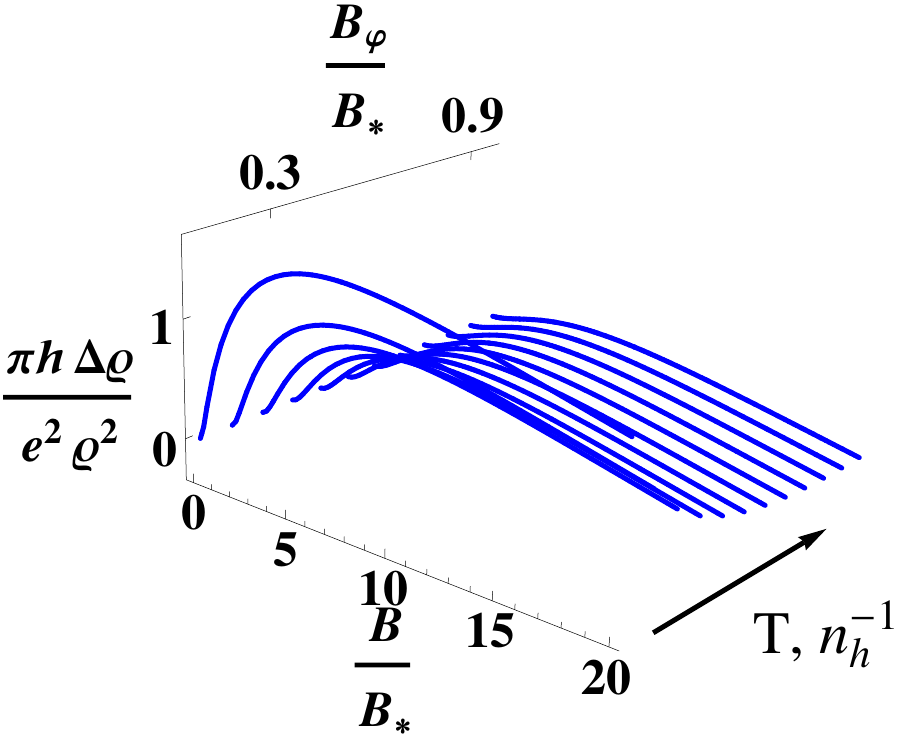}
 \caption{MR behaviour in regime A as a function of applied magnetic field, $B$, and phase-coherence lengths through $B_{\varphi}\propto \ell_{\varphi}^{-2}$. Left (in red): No inter-valley scattering, $\tau_{is}\rightarrow\infty$. WL saturates at temperatures/hole densities such that $\tau_{\varphi}\sim\tau_{*}$. Right (in blue): Inter-valley scattering is present, $\tau_{is}\sim\tau_{*}$. MR displays a crossover from WL to WAL upon increasing hole densities (decreasing $B_{\varphi}$).}
\label{fig:mag_2}
\end{figure}

\subsection*{Regime B: $\boldsymbol{\varepsilon_F\gtrsim\lambda}$ (valence band)}

When some minority spin carriers are present on top of majority spin Fermi seas in both valleys both intra-valley spin-flip and inter-valley spin-conserving scattering are permited for carriers at the Fermi level. Those are characterized by the intervalley, $\tau_{v}^{-1}$, and spin-flip intravalley, $\tau_{sf}^{-1}$, scattering rates, 
\begin{gather}
\tau_{v/sf}^{-1}=\frac{4\pi\nu }{\hbar}\left[\Upsilon_{v/sf}+ p_F^2\Xi_{v/sf} \right],
\label{v-sf}
\end{gather}
parameterised by the correlation functions,
\begin{gather}
\left\langle u_{sf}^i\left(\mathbf{r}\right)u_{sf}^j\left(\mathbf{r}'\right)\right\rangle=
\Upsilon_{sf}\delta_{ij}\delta\left(\mathbf{r}-\mathbf{r}'\right), 
\nonumber\\
\left\langle w_{\alpha}^{i}\left(\mathbf{r}\right)w_{\beta}^{j}\left(\mathbf{r}'\right)\right\rangle=
\Xi_{sf}\delta_{\alpha\beta}\delta_{ij}\delta\left(\mathbf{r}-\mathbf{r}'\right),
\nonumber
\\
\left\langle u_{i}^i\left(\mathbf{r}\right)u_{i}^j\left(\mathbf{r}'\right)\right\rangle=
\Upsilon_{v}\delta_{ij}\delta\left(\mathbf{r}-\mathbf{r}'\right), 
\nonumber\\
\left\langle w_{z\alpha}^i\left(\mathbf{r}\right)w_{z\beta}^j\left(\mathbf{r}'\right)\right\rangle=
\Xi_{v}\delta_{\alpha\beta}\delta_{ij}\delta\left(\mathbf{r}-\mathbf{r}'\right).
\nonumber
\end{gather}
In the crystals with short-range (atomic) defects, these rates are independent of the carrier density, whereas in a clean smoothly bent layers they would linearly increase with the density of carriers. 

Firing up new scattering processes at the carrier density threshold $n_c=n_h(\varepsilon_F=\lambda)$ reduces both the mean free path of carriers and their spin-diffusion length (now, limited by the intravalley spin-flip). The reduction of the mean free path by additional scattering channels leads to a step change in the resistivity and spin relaxation upon the increase of the density across the threshold,
\begin{gather}
\rho(n_h) \approx \rho(n_c) +\frac{m}{e^2n_h}(\tau_{v}^{-1}+\tau_{sf}^{-1})\theta(n_h-n_c), \\
\nonumber
\end{gather}
followed by a gradual decrease of resistivity, with a slope determined by the mobilities of majority- and minority-spin carriers. 

An abrupt change also occurs in the behavior of the quantum correction to conductivity of defected TMDCs. In short, openning intervalley and spin-flip intravalley scattering channel, with finite rates already at $n_h\geq n_c$, drives the system deeper into the symplectic symmetry class. This statement is based on the diagramatic calculation, where now, similarly to graphene,\cite{McCannFalko2012} one can classify Cooperons as singlets and triplets in terms of both spin ($s$) and valley ($l$) indexes,\begin{align*}
C_{ss'}^{ll'}\equiv\frac{1}{4}\left[s_ys_s\right]_{\alpha\beta}
\left[\tau_x\tau_l\right]^{ab}
C_{\alpha\beta\alpha'\beta'}^{aba'b'}
\left[s_{s'}s_y\right]_{\beta'\alpha'}
\left[\tau_{l'}\tau_x\right]^{b'a'}.
\end{align*}
The Bethe-Salpeter equations for the Cooperons can be written in a
compact way as\begin{align*}
C_{s_1s_2}^{l_1l_2}\left(\mathbf{Q},\omega\right)=W_{s_1s_2}^{l_1l_2}+\sum_{s,s'}\sum_{l,l'} W_{s_1s'}^{l_1l'}C_{ss_2}^{ll_2}\left(\mathbf{Q},\omega\right)\Pi_{ss'}^{ll'}\left(\mathbf{Q},\omega\right),
\end{align*}
where now the polarization operator reads
\begin{widetext}
\begin{align}
\Pi_{ss'}^{ll'}\left(\mathbf{Q},\omega\right)\equiv\frac{1}{4}\int \frac{d^2\mathbf{p}}{\left(2\pi\hbar\right)^2} \mbox{Tr}\left[\left(\tau_l\tau_x\right)\otimes\left(s_ss_y\right) \left(\hat{G}^R\left(\mathbf{p},\hbar\omega+\varepsilon_F\right)\right)^{T}
\left(\tau_x\tau_{l'}\right)\otimes\left(s_ys_{s'}\right)
\hat{G}^A\left(\mathbf{Q}-\mathbf{p},\varepsilon_F\right)\right],
\end{align}
\begin{center}
\begin{table}
\begin{tabular}{|c||c|}
\hline
Relaxation gaps&Relaxation rates\\
\hline
\hline
$\Gamma_0^0=0$&$\tau^{-1}=\tau_0^{-1}+\tau_{g}^{-1}+
\tau_{v}^{-1}+\tau_{sf}^{-1}+\tau_{is}^{-1}$\\
$\Gamma_x^0=\Gamma_y^0=\tau^{-1}_{\lambda}+2\tau^{-1}_{gz}+\tau^{-1}_{v}-
\tau^{-1}_{\gamma_v}
+\tau^{-1}_{sf}+\tau^{-1}_{is}$&\\
$\Gamma_z^0=2\tau_{sf}^{-1}+2\tau_{is}^{-1}$&$\tau_{gz}^{-1}=\frac{2\pi\nu p_F^2\Theta_{gz}}{\hbar}$\\
$\Gamma_0^x=\Gamma_0^y=\tau_{\lambda}^{-1}+\tau_{**}^{-1}+\tau_{sf}^{-1}+
\tau_{\gamma_{sf}}^{-1}
+\tau_{v}^{-1}+\tau_{is}^{-1}$
&\\
$\Gamma_x^x=\Gamma_x^y=\Gamma_y^x=\Gamma_y^y=\tau_{*}^{-1}+\tau_{v}^{-1}+
\tau_{sf}^{-1}+\tau_{is}^{-1}$&
$\tau_{**}^{-1}=\tau_{*}^{-1}-2\tau_{gz}^{-1}$\\
$\Gamma_z^x=\Gamma_z^y=\tau_{\lambda}^{-1}+\tau_{**}^{-1}+\tau_{sf}^{-1}-
\tau_{\gamma_{sf}}^{-1}+
\tau_{v}^{-1}+\tau_{is}^{-1}$&\\
$\Gamma_0^z=2\tau_{v}^{-1}+2\tau_{is}^{-1}$&
$\tau_{\gamma_{sf}}^{-1}=\frac{4\pi\nu }{\hbar}\left[\Upsilon_{sf}-p_F^2\Xi_{sf}\right]$\\
$\Gamma_x^z=\Gamma_y^z=\tau_{\lambda}^{-1}+2\tau_{gz}^{-1}+\tau_{v}^{-1}-
\tau_{\gamma_v}^{-1}
+\tau_{sf}^{-1}+\tau_{is}^{-1}$&\\
$\Gamma_z^z=2\tau_{v}^{-1}+2\tau_{sf}^{-1}$&
$\tau_{\gamma_{v}}^{-1}=\frac{4\pi\nu }{\hbar}\left[\Upsilon_{v}-p_F^2\Xi_{v}\right]$\\
\hline
\end{tabular}
\caption{Relation between Cooperon relaxation gaps and scattering rates in regimes B and C.}
\label{tab:gapsb}
\end{table}
\end{center}
\end{widetext}
and the disorder-averaged Green operators are
\begin{align}
\hat{G}^{R,A}\left(\mathbf{p},\omega\right)=\frac{\left(\hbar\omega-\varepsilon_{\mathbf{p}}\pm i\frac{\hbar}{2\tau}\right)+\frac{\lambda}{2}\tau_z\otimes s_z}
{\left(\hbar\omega-\varepsilon_{\mathbf{p}}\pm i\frac{\hbar}{2\tau}\right)^2-\frac{\lambda^2}{4}}.
\end{align}
Hence, the WL correction to conductivity are given by 16 
Cooperon modes ($\tilde{C}_s^{l}\equiv \frac{2\pi\nu\tau^2}{\hbar}C_{ss}^{ll}$),
\begin{gather}
\delta g_{ii}=-\frac{2e^2D}{\pi\hbar}\int
\frac{d^2\mathbf{Q}}{\left(2\pi\right)^2}\sum_{s,l}c_sc^l\tilde{C}_s^l\left(\mathbf{Q}\right),
\mbox{ with}
\nonumber\\
c_{0,x,y,z}=-1,+1,+1,+1, \nonumber\\c^{0,x,y,z}=+1,+1,+1,-1.
\label{eq:WL2}
\end{gather}

As before, we take the low momentum and frequency expansion of the polarization operator assuming the diffusive approximation, $\tau D\left|\mathbf{Q}\right|^2,\tau\omega\ll 1$. Two groups of 4 Cooperons corresponding to singlet and triplet combinations built separately of two Kramers doublets ($K_+,\uparrow$; $K_-,\downarrow$) and ($K_+,\downarrow$; $K_-,\uparrow$) are solutions of the diffusion-relaxation kernels\begin{align}
\left[-D\nabla^2-i\omega+\Gamma_s^{l}\right]\tilde{C}_s^{l}\left(\mathbf{r}-\mathbf{r}',\omega\right)=\delta\left(\mathbf{r}-\mathbf{r}'\right),
\end{align}
whereas the cross-doublet Cooperons do not contribute in the present regime due to the mismatch between the Fermi surfaces corresponding to different spin polarizations at each valley. The WL correction to conductivity and MR in this regime reads in general
\begin{widetext}
\begin{gather}
\delta g=\frac{e^2}{\pi h}\times\left[\ln\left(\frac{\tau^{-1}}{\tau_{\varphi}^{-1}}\right)
+\ln\left(\frac{\tau^{-1}}{\tau_{\varphi}^{-1}+\Gamma_z^z}\right)-
\ln\left(\frac{\tau^{-1}}{\tau_{\varphi}^{-1}+\Gamma_z^0}\right)-
\ln\left(\frac{\tau^{-1}}{\tau_{\varphi}^{-1}+\Gamma_0^z}\right)-
4\ln\left(\frac{\tau^{-1}}{\tau_{\varphi}^{-1}+\Gamma_x^x}\right)
\right],
\nonumber\\
\frac{\Delta\rho\left(B\right)}{\rho^2}=\frac{e^2}{\pi h}\left[F\left(\frac{B}{B_{\varphi}}\right)
+F\left(\frac{B}{B_{\varphi}+B_z^z}\right)
-F\left(\frac{B}{B_{\varphi}+B_z^0}\right)
-F\left(\frac{B}{B_{\varphi}+B_0^z}\right)
-4F\left(\frac{B}{B_{\varphi}+B_x^x}\right)\right],
\end{gather}
\end{widetext}
where the expressions for the relaxation gaps are summarized in Tab.~\ref{tab:gapsb} and we have introduced $B_{s}^l=\frac{\hbar\Gamma_s^l}{4eD}$.

These expressions can be further simplified if we neglect $\tau_{*}^{-1}$ and $\tau_{is}^{-1}$ in the expression for $\Gamma_x^x$, $\Gamma_0^z$, and $\Gamma_z^0$, which are assume to be smaller due to their dependence on carrier concentration. Then, we have\begin{gather}
\delta g=\frac{e^2}{\pi h}\left[
\ln\frac{(\tau_{\varphi}^{-1}+2\tau_{sf}^{-1})(\tau_{\varphi}^{-1}+2\tau_{v}^{-1})}{\tau_{\varphi}^{-1}(\tau_{\varphi}^{-1}+2\tau_{v}^{-1}+2\tau_{sf}^{-1})}\right.
\nonumber\\
\left. -4\ln\left(\frac{\tau^{-1}}{\tau_{\varphi}^{-1}+\tau_{v}^{-1}+\tau_{sf}^{-1}}\right)\right];
\label{WLb}
\\
\frac{\Delta\rho\left(B\right)}{\rho^2}=\frac{e^2}{\pi h}F\left(\frac{B}{B_{\varphi}}\right)
\nonumber\\
-\frac{e^2}{96\pi h}\left[\frac{15B^2}{(B_v+B_{sf})^2}+\frac{B^2}{B_v^2}+\frac{B^2}{B_{sf}^2}  \right],
\nonumber
\end{gather}
where we have expanded all the $F\left(z\right)$ functions to the lowest order excepting the first term and $B_{\alpha}=\hbar/(4e\tau_{\alpha}D)$. Note that in this regime intravalley spin-flip processes do not lead immediately to WAL. This happens because the 4$^{th}$ term in the 1$^{st}$ line of Eq.~\eqref{eq:dis} looks like intravalley magnetic disorder which suppresses WAL effect coming from intravalley spin-flip scattering, similarly to how trigonal warping and Berry curvature (accounted by $\tau_*^{-1}$ in Eqs.~\eqref{crossover-g}-\eqref{crossover-MR}) suppress WL in the transport regime A. 

\subsection*{Regime C: $\boldsymbol{\varepsilon_F \gg \lambda}$ (conduction band)}

In this case, the crossover between WL and WAL behaviour of magnetoresistance takes the most complicated form, especially when $\lambda\tau<\hbar$, since one has to take into account all 16 Cooperons built using valley-spin quartet ($K_+,\uparrow; K_-,\downarrow; K_+,\downarrow; K_-,\uparrow$). The intrinsic SO splitting acts as an effective Zeeman coupling making the electron spin precess around the axis perpendicular to the crystal plane in the opposite direction for electrons in the opposite valleys. SO splitting and spin-dependent disorder determine the rate,
\begin{align}
\tau_{\lambda}^{-1}=\frac{\lambda^2\tau}{\hbar^2}+\frac{4\pi\nu\Omega_z}{\hbar},
\end{align}
at which the 8 Cooperon modes that we neglected previously, $\tilde{C}_{x,y}^{0,z}$ and $\tilde{C}_{0,z}^{x,y}$, decay. The SO splitting also couples them by precession with the rate $\omega_{\lambda}\equiv\lambda/\hbar$, similarly to how real Zeeman coupling mixes singlet/triplet Cooperon modes in a simple disordered metal.\cite{japos} These modes satisfy the matrix equations (for the former)\begin{align}
\left(\begin{array}{cc}
\Pi+\Gamma_{x}^{0(z)} &-\omega_{\lambda}\\
\omega_{\lambda}&\Pi+\Gamma_{y}^{z(0)}
\end{array}\right)\cdot\left(\begin{array}{cc}
\tilde{C}_{xx}^{00(zz)}&\tilde{C}_{xy}^{0z(z0)}\\
\tilde{C}_{yx}^{z0(0z)}&\tilde{C}_{yy}^{zz(00)}
\end{array}\right)=\mathcal{I},
\end{align}
where we have written $\Pi=D\left|\mathbf{Q}\right|^2-i\omega$ for simplicity. After matrix inversion we have
\begin{gather}
\tilde{C}_x^{0(z)}=\frac{\Pi+\Gamma_{y}^{z(0)}}{\left(\Pi+\Gamma_{x}^{0(z)}\right)\left(\Pi+\Gamma_{y}^{z(0)}\right)+\omega_{\lambda}^2},\nonumber\\
\tilde{C}_y^{z(0)}=\frac{\Pi+\Gamma_{x}^{0(z)}}{\left(\Pi+\Gamma_{y}^{z(0)}\right)\left(\Pi+\Gamma_{x}^{0(z)}\right)+\omega_{\lambda}^2}.
\end{gather}
For $\tilde{C}_{0,z}^{x,y}$ we obtain the same swaping the spin and valley indices. Then, by introducing the coefficients\begin{gather*}
\gamma_{v}\equiv\frac{\tau_{\gamma_v}^{-1}}{\sqrt{\tau_{\gamma_v}^{-2}-\omega_{\lambda}^2}},\nonumber\\
\gamma_{sf}\equiv\frac{\tau_{\gamma_{sf}}^{-1}}{\sqrt{\tau_{\gamma_{sf}}^{-2}-\omega_{\lambda}^2}},
\end{gather*}
where the rates $\tau_{\gamma_{v}}^{-1}$, $\tau_{\gamma_{sf}}^{-1}$ are defined in the second coluumn of Tab.~\ref{tab:gapsb}, the WL correction to conductivity and MR can be written as
\begin{widetext}
\begin{gather}
\delta g=\frac{e^2}{\pi h}\times\left\{\ln\left(\frac{\tau^{-1}}{\tau_{\varphi}^{-1}}\right)
+\ln\left(\frac{\tau^{-1}}{\tau_{\varphi}^{-1}+\Gamma_z^z}\right)-
\ln\left(\frac{\tau^{-1}}{\tau_{\varphi}^{-1}+\Gamma_z^0}\right)-
\ln\left(\frac{\tau^{-1}}{\tau_{\varphi}^{-1}+\Gamma_0^z}\right)-
4\ln\left(\frac{\tau^{-1}}{\tau_{\varphi}^{-1}+\Gamma_x^x}\right)+
\right.\nonumber\\\left.
+2\gamma_v\left[\ln\left(\frac{\tau_{\gamma_v}^{-1}+\tau^{-1}\gamma_v}{\tau_{\gamma_v}^{-1}-\tau^{-1}\gamma_v}\right)-
\ln\left(\frac{\tau_{\gamma_v}^{-1}+\left(\tau_{\varphi}^{-1}+\Gamma_x^0/2+
\Gamma_x^z/2\right)\gamma_v}{\tau_{\gamma_v}^{-1}-\left(\tau_{\varphi}^{-1}+\Gamma_x^0/2+
\Gamma_x^z/2\right)\gamma_v}\right)\right]+
\right.\nonumber\\\left.
+2\gamma_{sf}\left[\ln\left(\frac{\tau_{\gamma_{sf}}^{-1}+\tau^{-1}\gamma_s}{\tau_{\gamma_{sf}}^{-1}-\tau^{-1}\gamma_s}\right)-
\ln\left(\frac{\tau_{\gamma_{sf}}^{-1}+\left(\tau_{\varphi}^{-1}+\Gamma_0^x/2+
\Gamma_z^x/2\right)\gamma_s}{\tau_{\gamma_{sf}}^{-1}-\left(\tau_{\varphi}^{-1}+\Gamma_0^z/2+
\Gamma_z^x/2\right)\gamma_s}\right)\right]
\right\},
\nonumber\\
\frac{\Delta\rho\left(B\right)}{\rho^2}=\frac{e^2}{\pi h}\left[F\left(\frac{B}{B_{\varphi}}\right)
+F\left(\frac{B}{B_{\varphi}+B_z^z}\right)
-F\left(\frac{B}{B_{\varphi}+B_z^0}\right)
-F\left(\frac{B}{B_{\varphi}+B_0^z}\right)
-4F\left(\frac{B}{B_{\varphi}+B_x^x}\right)+
\right.\nonumber\\\left.
+2\gamma_vG\left(\frac{B}{B_{\varphi}+\frac{B_x^0+B_x^z}{2}},\frac{\gamma_vB}{B_{\gamma_v}}\right)
+2\gamma_{sf}G\left(\frac{B}{B_{\varphi}+\frac{B_0^x+B_z^x}{2}},\frac{\gamma_{sf}B}{B_{\gamma_{sf}}}\right)
\right],
\end{gather}
\end{widetext}
where
\begin{align*}
G\left(z_1,z_2\right)
\equiv
\sum_{\pm}\left[
\pm\psi\left(\frac{1}{2}+\frac{1}{z_1}\pm\frac{1}{z_2}\right)\mp\ln\left(
\frac{1}{z_1}\pm\frac{1}{z_2}\right)
\right].
\end{align*}

We analyze the quantum transport behaviour in TMDCs deduced from these formulae in two extreme situations depending on their crystalline quality: (i) material where scattering is dominated by lattice defects and (ii) defect-free TMDC.

\subsubsection{Lattice-disordered TMDCs}

We take into account only such disorder that leads to finite scattering rates $\tau^{-1}_{0,v,sf}$ for electrons at the edge of conduction band, leading to \begin{align*} \gamma_{v,sf}=\frac{1}{\sqrt{1-\frac{\lambda^2\tau_{v,sf}^2}{\hbar^2}}}.
\end{align*} In this case, MR has a distinct WAL form, extrapolated from the WAL behaviour in the regime B,
\begin{gather}
\frac{\Delta\rho\left(B\right)}{\rho^2}
=\frac{e^2}{\pi h}F\left(\frac{B}{B_{\varphi}}\right)
\label{eq:magtop}\\
-\frac{e^2}{96\pi h}\left[\frac{15B^2}{(B_v+B_{sf})^2}+\frac{B^2}{B_v^2}+\frac{B^2}{B_{sf}^2}  \right.
\nonumber\\
\left. +\sum_{\alpha=v,sf}
\frac{32 B_{\alpha}\left(B_{\lambda}+B_v+B_{sf}\right)B^2}{\left[\tilde{B}_{\lambda}^2+\left(B_{\lambda}+B_{sf}+B_v\right)^2-B_{\alpha}^2\right]^2}  \right],
\nonumber
\end{gather}
where we have expanded $F(z)$ and $G(z)$ to the lowest order in $z$ and $B_{\alpha}=\hbar/(4e\tau_{\alpha}D)$, $\tilde{B}_{\lambda}=\lambda/(4eD)$.

\subsubsection{TMDCs free of atomic defects}

In a defect-free 2D crystal with scattering produced by remote charges in the substrate or smooth lattice deformations, electrons diffuse conserving their valley state ($\tau_{v,is}\rightarrow\infty$). Then, spin-diffusion lengths,
\begin{align}
L_s^{(c)}\sim\frac{\hbar p_F}{\lambda m_*}\times
\max \left[
\frac{\mathcal{L}}{\sqrt{2\left\langle h^2 \right\rangle}},
\sqrt{\frac{2\pi\kappa}{K_B T}}
\right] \propto \sqrt{n_e},
\end{align}
are assumed to be limited by either the characteristic height $\sqrt{\left\langle h^2 \right\rangle}$ of static wrinkles of lateral size $\mathcal{L}$, or temperature in the case of flexural vibration modes, where $\kappa$ is bending stiffness of the 2D crystal.\cite{ochoa_etal,ochoa_roldan}

As to the quantum transport, spurious time-inversion asymmetry for the intravalley electron propagation caused by SO coupling, Berry phase and pseudomagnetic field due to the deformations, suppress the interference correction to conductivity. As a result, MR in such high-quality 2D material would have a form of a suppressed WL effect, 
\begin{gather}
\frac{\Delta\rho\left(B\right)}{\rho^2}=-\frac{2e^2}{\pi h}\left\{
2F\left(\frac{B}{B_{\varphi}+B_{sf}}\right)+
\frac{1}{\sqrt{1-\frac{\lambda^2\tau_{sf}^2}{\hbar^2}}}\times
\right.\nonumber\\
\left. \times G\left(\frac{B}{B_{\varphi}+
B_{\lambda}+B_{sf}},\frac{B}{B_{sf}\sqrt{1-\frac{\lambda^2\tau_{sf}^2}{\hbar^2}}}\right)\right\}.
\label{eq:magbottom}
\end{gather}

\section{Discussion}

For n-doped TMDCs (regime C) it is interesting to discuss the extreme of samples free of atomic defects, where the expression for the MR adopts its easiest form, Eq.~\eqref{eq:magbottom}. We see that the suppression of WL is governed by the ratio between the spin-relaxation rate, $\tau_{sf}^{-1}$, and the characteristic precession frequency defined by the SO splitting, $\omega_{\lambda}$. Taking MoS$_2$ as a reference, we have $\tau_{sf}\sim 1$ ns,\cite{ochoa_etal,ochoa_roldan} which is compatible with optical experiments,\cite{Mak_etal} whereas the precession frequencies are of the order of THzs, leading to $\tau_{sf}\omega_{\lambda}\sim10^3$. Thus, WL behaviour is expected, eventually suppressed by warping or Berry phase effects. Hence, if an experiment on n-doped MoS$_2$ displayed WAL behaviour, this would inmediately point at the presence of short-range disorder (such as vacancies in the chalcogen atoms layer, which break the $z\rightarrow-z$ symmetry of the system) which scatters between valleys simultaneously flipping spins.

For p-doped TMDCs spin relaxation has a rate linear in the carrier density for $\varepsilon_F(n_h)<\lambda$, which leads to a density-independent spin-diffusion length, and that their MR displays a crossover from WL to WAL behavior upon the increase in the concentration of holes. At the threshold density, $n_c$ of the population of minority spin states in each valley,  $\varepsilon_F(n_c)=\lambda$, resistivity and spin relaxation rate of holes udergo a step-like increase, whereas the quantum correction to conductivity remains of a WAL type.

To complete this dicussion, we also consider a generally ignored but possible ocurrence of the $\Gamma$-point band edge in some TMDCs. To the lowest orders in momentum, the $\mathbf{k}\cdot\mathbf{p}$ Hamiltonian, including SO terms, reads
\begin{align}
\mathcal{H}=\frac{\left|\mathbf{p}\right|^2}{2m^*}+
\alpha\left(p_x^3-3p_xp_y^2\right)s_z.
\label{HGamma}
\end{align}
The SO parameter can be roughly estimated as $\alpha\approx\lambda/\left(\hbar^3|K_{\pm}|^3\right)=\left(\frac{\sqrt{3}a}{4\pi\hbar}\right)^3\lambda$. For a finite concentration of holes around $\Gamma$, $n_{\Gamma}$, we have\begin{align*}
\frac{\alpha p_F^3}{\hbar/\tau}\sim \ell a^3n_c^2\times\left(\frac{n_{\Gamma}}{n_c}\right)^{\frac{3}{2}}\ll 1.
\end{align*}
Hence, the spin splitting of electron states near $\Gamma$-point plays no role, whereas $z\rightarrow-z$ symmetry breaking flexural deformations and substrate induced asymmetry would lead to the typical Bychkov-Rashba SO effects,\cite{BR} Dyakonov-Perel\cite{DP1,DP2} spin relaxation and WL-WAL crossover, as in GaAs/AlGaAs heterostructures.\cite{aleiner_falko,Miller_etal,Cremers_etal} This should be contrasted with suppressed WL behaviour in high-quality, low-to-medium p-doped samples characteristic for $K_{\pm}$ points band edges. With the references to Eqs.~\eqref{crossover-g}-\eqref{crossover-MR}, we suggest that if fitting of experimentally measured magnetoresistance returned $\tau_{*}$ such that $\tau\ll\tau_*\ll\tau_{\varphi}$, this would give a distinct quantitative proof for the multi-valley nature of the valence band edge.

\section{Acknowledgements}

The authors thank I. Aleiner and A. Morpurgo for useful discussions. This work was supported
by CSIC JAE-Pre grant, EC FP7 Graphene Flagship project CNECT-ICT-604391, ERC Synergy Grant Hetero2D, the Royal
Society Wolfson Research Merit Award, Marie-Curie-ITN 607904-13 Spinograph, and ERC Advanced Grant 290846.

\appendix

\section{$\mathbf{k}\cdot\mathbf{p}$ theory for \\electrons and holes in TMDCs}
\label{sec:AppA}

In this appendix we deduce the band Hamiltonian of Eq.~\eqref{Ham} from a $\mathbf{k}\cdot\mathbf{p}$ theory describing lowest conduction and valence bands.\cite{two_band,2DBands5} These are dominated by $d$ orbitals from the $M$ atoms, $d_{3z^2-r^2}$ and $\left(d_{x^2-y^2}\pm id_{xy}\right)$ respectively. Instead of dealing with degenerate states at $\mathbf{K}_{\pm}$ points one can triple the unit cell in such a way that the old $K_{\pm}$ points are now equivalent to the $\Gamma$ point of the folded Brillouin zone. From the point of view of the lattice symmetries, this means that the two elementary translations ($\mathbf{t}_{\mathbf{a}_1}$,$\mathbf{t}_{\mathbf{a}_2}$) are factorized out of the translation group and added to the point group $D_{3h}$,
which becomes $D_{3h}''=D_{3h}+\mathbf{t}_{\mathbf{a}_1}\times D_{3h}+\mathbf{t}_{\mathbf{a}_2}\times D_{3h}$. The character table of this group is shown in Tab.~1. $D_{3h}''$ contains 24 new elements and 6 additional conjugacy classes, which leads to 6 new 2-dimensional irreducible representations (denoted by $E_{1,2,3}'$ and $E_{1,2,3}''$), the valley off-diagonal representations.

\begin{widetext}
\begin{center}
\begin{table}
\begin{tabular}{|c||c|c|c|c|c|c|c|c|c|c|c|c|}
\hline
$D_{3h}''$&$E$&$2T$&$\sigma_h$&$2T\sigma_h$
&$2C_{3}$&$2TC_3$&$2TC_3^2$&$2S_3$&$2TS_3$&$2TS_3^2$&$9TC_2'$&$9T\sigma_v$\\
\hline
\hline
$A_1'$&1&1&1&1&1&1&1&1&1&1&1&1\\
\hline
$A_2'$&1&1&1&1&1&1&1&1&1&1&-1&-1\\
\hline
$A_1''$&1&1&-1&-1&1&1&1&-1&-1&-1&1&-1\\
\hline
$A_2''$&1&1&-1&-1&1&1&1&-1&-1&-1&-1&1\\
\hline
$E'$&2&2&2&2&-1&-1&-1&-1&-1&-1&0&0\\
\hline
$E''$&2&2&-2&-2&-1&-1&-1&1&1&1&0&0\\
\hline
$E_1'$&2&-1&2&-1&2&-1&-1&2&-1&-1&0&0\\
\hline
$E_1''$&2&-1&-2&1&2&-1&-1&-2&1&1&0&0\\
\hline
$E_2'$&2&-1&2&-1&-1&2&-1&-1&2&-1&0&0\\
\hline
$E_2''$&2&-1&-2&1&-1&2&-1&1&-2&1&0&0\\
\hline
$E_3'$&2&-1&2&-1&-1&-1&2&-1&-1&2&0&0\\
\hline
$E_3''$&2&-1&-2&1&-1&-1&2&1&1&-2&0&0\\
\hline
\end{tabular}
\caption{Character table of $D_{3h}''$.}
\label{tab:character}
\end{table}
\end{center}
\end{widetext}

The symmetry properties of Bloch wave functions at the Brillouin zone corners are summarized in Tab.~\ref{tab:wf}, which gives the suitable combination of atomic orbitals and the associated irreducible representation of $D_{3h}''$. In the case of $X$ atoms, both bonding (b) and anti-bonding (ab) combinations of orbitals from the bottom and top layers are considered. The second and third column contain the phases picked up by the wave function at each valley when a $2\pi/3$ rotation or a mirror reflection is performed. We consider the space of 4-vectors $\sim \left(E_2',E_1'\right)$ whose entries represent the projection of the Bloch wave function at conduction and valence states at the Brillouin zone corners. In order to construct the effective $\mathbf{k}\cdot\mathbf{p}$ Hamiltonian acting on this subspace we must consider the possible 16 hermitian operators, whose reduction in terms of irreducible representations of $D_{3h}''$ is inferred from:\begin{align*}
\left(E_2',E_1'\right)\times\left(E_2',E_1'\right)\sim2A_1'+2A_2'+2E'+E_1'+E_2'+2E_3'.
\end{align*}
This space of electronic operators can be constructed from two conmutating Pauli algebras $\Sigma_i$, $\Lambda_i$. The definitions are summarized in Tab.~\ref{tab:electronic_operators2}. The basis is $\left(\psi_{c+},\psi_{v+},\psi_{v-},-\psi_{c-}\right)$, where $\psi_{c,v\pm}$ represents the wave function of the conduction or valence state at $K_{\pm}$ points, in such a way that time reversal operation (including spin) reads $is_y\Lambda_y\Sigma_y\mathcal{K}$. The operators $\Lambda_i$, $\Sigma_i$ and all their combinations are $4\times4$ matrices, which in this basis can be written as\begin{eqnarray*}
\Sigma_{x,y,z}=\tau_0\otimes\sigma_{x,y,z},\nonumber\\
\Lambda_{x,y,z}=\tau_{x,y,z}\otimes\sigma_{0},
\end{eqnarray*}where $\tau_{i}$ and $\sigma_i$ are Pauli matrices that act in valley and conduction/valence subspaces.

The Hamiltonian up to second order in $\mathbf{p}$ reads\cite{foot_gap}\begin{gather}
\mathcal{H}=\gamma\mathbf{p}\cdot\boldsymbol{\Sigma}+\frac{\Delta}{2}\Lambda_z\Sigma_z+\frac{\alpha+\beta}{2}\mathcal{I}|\mathbf{p}|^2+
\nonumber\\
+\frac{\alpha-\beta}{2}\Lambda_z\Sigma_z|\mathbf{p}|^2+\kappa\left[\left(p_x^2-p_y^2\right)\Lambda_z\Sigma_x-2p_xp_y\Lambda_z\Sigma_y\right],
\label{eq:model}
\end{gather}
which corresponds to the Hamiltonian in Eqs.~(2a)-(2d) of Ref.~\onlinecite{2DBands5}. Microscopically, the linear term in $\mathbf{p}$ comes from the strong hybridization between conduction and valence band states away from $K_{\pm}$ points, both dominated by orbitals localized in the metal transition atoms. Such hybridization is responsible for the non-zero Berry curvature of the bands,\cite{berry}
\begin{align}
\Omega_{c,v}^{\tau}\left(\mathbf{p}\right)\approx\mp\frac{2\tau\gamma^2\left[\Delta-
\left(\alpha-\beta\right)|\mathbf{p}|^2\right]}{\left[\left(\Delta+\left(\alpha-\beta\right)^2|\mathbf{p}|^2\right)+4\gamma^2|\mathbf{p}|^2\right]^{3/2}}.\end{align} The different orbital composition of conduction and valence bands introduce certain electron-hole asymmetry, and it is also responsible for the distinct strengths of the intrinsic spin-orbit splitting of the bands. We aslo include trigonal warping effects in the bands through the last term of the Hamiltonian. The isoenergy contours around $K_{\pm}$ points deduced from this model are shown in Fig.~\ref{fig:contours}.

We project the Hamiltonian of Eq.~\eqref{eq:model} onto a single band by a Schrieffer-Wolf transformation.\citep{SW}  The Hamiltonian can be written in the block form as
\begin{align}
\mathcal{H}=\left(\begin{array}{cc}
\mathcal{H}_{c}^{(0)} & V \\
V^{\dagger} & \mathcal{H}_{v}^{(0)}
\end{array}\right).
\end{align}
We take the Green function $\mathcal{G}=\left(\epsilon-\mathcal{H}\right)^{-1}$, then evaluate the block $\mathcal{G}_{c,v}$ associated to the conductiuon/valence band, and use it in order to identify the effective Hamiltonian near the band edge. If we define $\mathcal{G}_{c,v}^{(0)}=\left(\epsilon-\mathcal{H}_{c,v}^{(0)}\right)^{-1}$, then we can write\begin{align}
\left(\begin{array}{cc}
\mathcal{G}_{c} & \mathcal{G}_{cv} \\
\mathcal{G}_{vc} & \mathcal{G}_{v}
\end{array}\right)=\left(\begin{array}{cc}
\left(\mathcal{G}_{c}^{(0)}\right)^{-1} & V\\
V^{\dagger}& \left(\mathcal{G}_{v}^{(0)}\right)^{-1}
\end{array}\right)^{-1}.
\end{align}
For the conduction band, we obtain $\mathcal{G}_{c}=\left[\left(\mathcal{G}_{c}^{(0)}\right)^{-1}-V
\mathcal{G}_{v}^{(0)}V^{\dagger}\right]^{-1}$,
so $\epsilon-\mathcal{G}_{c}^{-1}=\mathcal{H}_{c}+V
\mathcal{G}_{v}^{(0)}V^{\dagger}$.
At the bottom of the band ($\epsilon\approx \frac{\Delta}{2}$) the effective Hamiltonian to the lowest order in $\Delta^{-1}$ reads
\begin{align}
\mathcal{H}_{c}\approx \mathcal{H}_{c}^{(0)}+\frac{V\left(\begin{array}{cc}
0&0\\
0&1
\end{array}\right)V^{\dagger}}{\Delta}.
\label{eq:Heffc}
\end{align}
Similarly, for the valence band we have
\begin{align}
\mathcal{H}_{v}\approx \mathcal{H}_{v}^{(0)}-\frac{V^{\dagger}\left(\begin{array}{cc}
1&0\\
0&0
\end{array}\right)V}{\Delta}.
\label{eq:Heffv}
\end{align}
By this procedure we obtain the Hamiltonian in Eq.~\eqref{Ham} with\begin{gather}
m_{c,v}^*=\frac{1}{2\left(\alpha,\beta\pm\frac{\gamma^2}{\Delta}\right)},\nonumber\\
\mu_{c,v}=\pm\frac{2\gamma\kappa}{\Delta}.
\end{gather}

\begin{center}
\begin{table}
\begin{tabular}{|c|c|c||c|c|}
\hline
Irreps&$C_3$&$\sigma_h$&$M$ atom&$X$ atoms\\
\hline
\hline
$E_1'$&1&1&$\begin{array}{c}
\frac{1}{\sqrt{2}}\left(d_{x^2-y^2}\pm id_{xy}\right),\\
\frac{1}{\sqrt{2}}\left(p_x\mp ip_{y}\right)
\end{array}$
&$\frac{1}{\sqrt{2}}\left(p_x\pm ip_{y}\right)$ (b)\\
\hline
$E_1''$&1&-1&$\frac{1}{\sqrt{2}}\left(d_{xz}\mp id_{yz}\right)$&$\frac{1}{\sqrt{2}}\left(p_x\pm ip_{y}\right)$ (ab)\\
\hline
$E_2'$&$w^{\pm1}$&1&$d_{3z^2-r^2}$, $s$&$\frac{1}{\sqrt{2}}\left(p_x\mp ip_{y}\right)$ (b)\\
\hline
$E_3'$&$w^{\mp1}$&1&
$\begin{array}{c}
\frac{1}{\sqrt{2}}\left(d_{x^2-y^2}\mp id_{xy}\right),\\
\frac{1}{\sqrt{2}}\left(p_x\pm ip_{y}\right)
\end{array}$
&$p_z$ (ab), $s$ (b)\\
\hline
$E_2''$&$w^{\pm1}$&-1&$p_z$&$\frac{1}{\sqrt{2}}\left(p_x\mp ip_{y}\right)$ (ab)\\
\hline
$E_3''$&$w^{\mp1}$&-1&$\frac{1}{\sqrt{2}}\left(d_{xz}\pm id_{yz}\right)$&$p_z$ (b), $s$ (ab)\\
\hline
\end{tabular}
\caption{Classification of the Bloch wave functions at $K_{\pm}$ according to the irreducible representations of $D_{3h}''$. The sign $\pm$ corresponds to combinations of orbitals at $K_{\pm}$ points, and $w=e^{i\frac{2\pi}{3}}$.}
\label{tab:wf}
\end{table}
\end{center}

\begin{center}
\begin{table}
\begin{tabular}{|c||c|c|}
\hline
Irrep&$t\rightarrow -t$ even&$t\rightarrow -t$ odd\\
\hline
\hline
$A_1'$&$\mathcal{I}$, $\Lambda_z\Sigma_z$, $\Lambda_zs_z$, $\Sigma_z s_z$&\\
\hline
$A_2'$&&$\Sigma_z$, $\Lambda_z$, $s_z$\\
\hline
$A_1''$&$\Sigma_xs_x+\Sigma_ys_y$&\\
\hline
$A_2''$&$\Sigma_xs_y-\Sigma_ys_x$&\\
\hline
$E'$&$\left(\begin{array}{c}
-\Lambda_z\Sigma_y\\
\Lambda_z\Sigma_x
\end{array}\right)$, $\left(\begin{array}{c}
-s_z\Sigma_y\\
s_z\Sigma_x
\end{array}\right)$&$\left(\begin{array}{c}
\Sigma_x\\
\Sigma_y
\end{array}\right)$\\
\hline
$E''$&$\left(\begin{array}{c}
\Lambda_zs_x\\
\Lambda_zs_y
\end{array}\right)$, $\left(\begin{array}{c}
\Sigma_zs_x\\
\Sigma_zs_y
\end{array}\right)$, $\left(\begin{array}{c}
\Sigma_xs_y+\Sigma_ys_x\\
\Sigma_xs_x-\Sigma_ys_y
\end{array}\right)$&$\left(\begin{array}{c}
-s_y\\
s_x
\end{array}\right)$\\
\hline
$E_1'$&$\left(\begin{array}{c}
\Sigma_x\Lambda_x+\Sigma_y\Lambda_y\\
\Sigma_x\Lambda_y-\Sigma_y\Lambda_x
\end{array}\right)$&\\
\hline
$E_1''$&$\left(\begin{array}{c}
\Lambda_xs_y-\Lambda_ys_x\\
\Lambda_xs_x+\Lambda_y s_y
\end{array}\right)$&\\
\hline
$E_2'$&$\left(\begin{array}{c}
\Sigma_y\Lambda_y-\Sigma_x\Lambda_x\\
-\Sigma_x\Lambda_y-\Sigma_y\Lambda_x
\end{array}\right)$&\\
\hline
$E_2''$&$\left(\begin{array}{c}
\Lambda_xs_y+\Lambda_ys_x\\
\Lambda_ys_y-\Lambda_x s_x
\end{array}\right)$&\\
\hline
$E_3'$&$\left(\begin{array}{c}
-\Sigma_z\Lambda_y\\
\Sigma_z\Lambda_x
\end{array}\right)$, $\left(\begin{array}{c}
-s_z\Lambda_y\\
s_z\Lambda_x
\end{array}\right)$&$\left(\begin{array}{c}
\Lambda_x\\
\Lambda_y
\end{array}\right)$\\
\hline
\end{tabular}
\caption{Definitions of the electronic operators in the two bands effective model at $K_{\pm}$ points.}
\label{tab:electronic_operators2}
\end{table}
\end{center} 

\begin{figure}
\includegraphics[width=0.49\columnwidth]{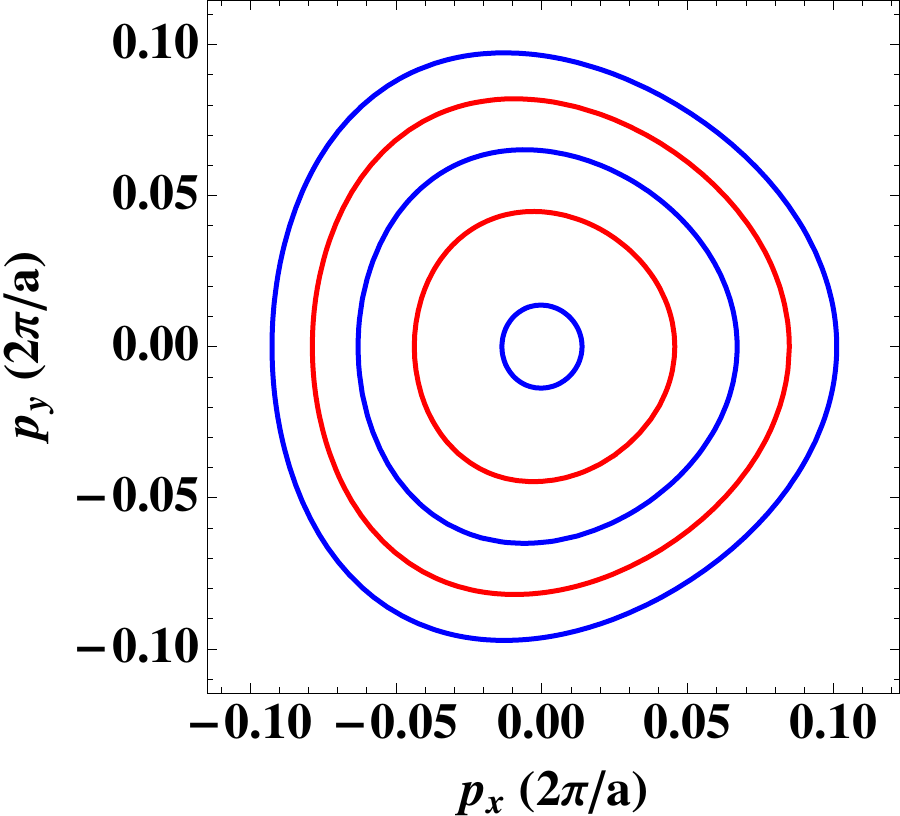}
\includegraphics[width=0.49\columnwidth]{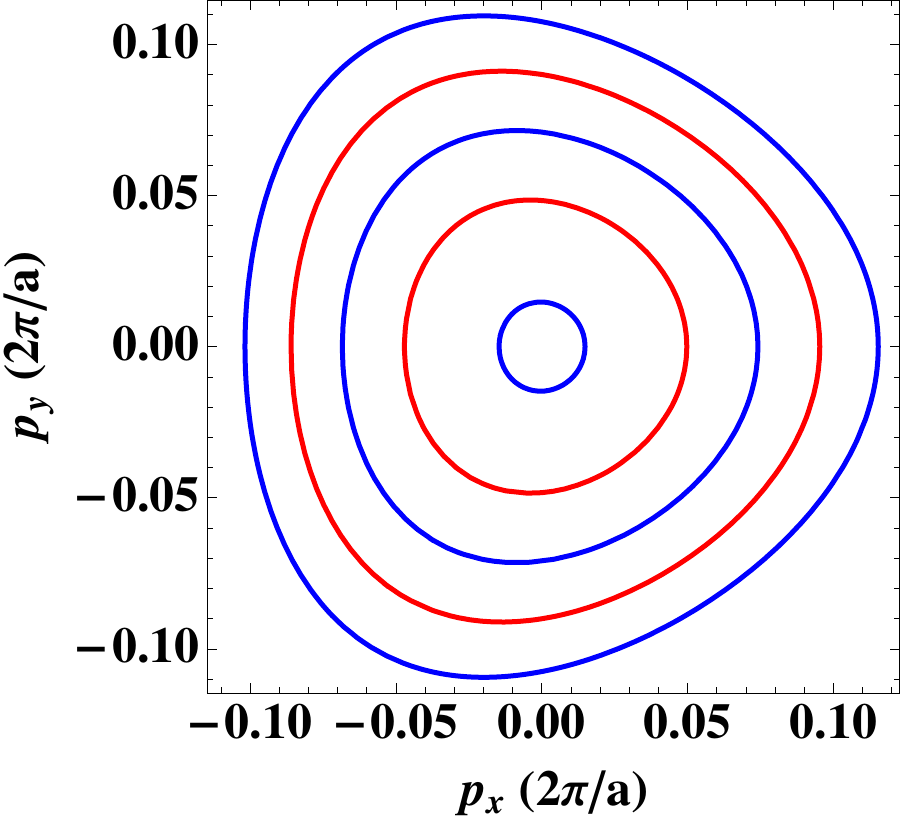}
 \caption{Isoenergy contours around $K_{\pm}$ points deduced from Eq.~\eqref{eq:model} for conduction (left) and valence (right) bands. We take $\gamma=3.82$ eV$\cdot$A, $\alpha=1.72$ eV$\cdot$A$^2$, $\beta=-0.13$ eV$\cdot$A$^2$, and $\kappa=-1.02$ eV$\cdot$A$^2$.\cite{2DBands5} The spin-orbit splitting is neglected.}
\label{fig:contours}
\end{figure}

\section{$\mathbf{k}\cdot\mathbf{p}$ theory for \\intra- and inter-valley disorder}
\label{sec:AppB}

In this appendix we deduce the form of the disorder potentials included in Eq.~\eqref{eq:dis}. Within the two-bands $\mathbf{k}\cdot\mathbf{p}$ theory, intra-valley disorder enters as \textit{scalar}, \textit{mass}, and \textit{gauge}-like potentials,\begin{align}
\delta\mathcal{H}\left(\mathbf{r}\right)=U\left(\mathbf{r}\right)\mathcal{I}+
M\left(\mathbf{r}\right)\Lambda_z\Sigma_z+\Lambda_z\boldsymbol{\Sigma}\cdot \mathbf{A}\left(\mathbf{r}\right).
\end{align}
When we project these terms onto a single band by the same procedure as before we obtain
\begin{align}
\delta\mathcal{H}_{c,v}=U_{c,v}\left(\mathbf{r}\right)\pm\frac{\gamma}{\Delta}
\left\{\mathbf{p},\mathbf{A}\left(\mathbf{r}\right)\right\}\tau_z,
\label{eq:disorder}
\end{align}
with $U_{c,v}\left(\mathbf{r}\right)=U\left(\mathbf{r}\right)\pm M\left(\mathbf{r}\right)+
\frac{\gamma}{\Delta}\left[\nabla\times\mathbf{A}\left(\mathbf{r}\right)\right]_z$. This corresponds to 1$^{st}$ and 3$^{rd}$ terms of Eq.~\eqref{eq:dis} of the main text with\begin{gather}
u_0\left(\mathbf{r}\right)=U\left(\mathbf{r}\right)\pm M\left(\mathbf{r}\right)+
\frac{\gamma}{\Delta}\left[\nabla\times\mathbf{A}\left(\mathbf{r}\right)\right]_z,\nonumber\\
\mathbf{a}_g\left(\mathbf{r}\right)=\pm\frac{\gamma}{\Delta}\mathbf{A}\left(\mathbf{r}\right).
\end{gather}

Inter-valley disorder can be incorporated following the same procedure. Within the two-bands model we have in general\begin{align}
V^{int}\left(\mathbf{r}\right)=\sum_{n=x,y,z}\sum_{l=x,y}V_{nl}\left(\mathbf{r}\right)
\Sigma_n\Lambda_l.
\label{eq:dis_v}
\end{align}
We can write down the spin-dependent disorder potentials in the same fashion, distinguishing even and odd terms under the reflection symmetry $z\rightarrow-z$ defined by the layer of transition metal atoms,\begin{gather}
V^{e}\left(\mathbf{r}\right)=
s_z\left[\sum_{n=x,y,z}\mathcal{U}_n^{e}\left(\mathbf{r}\right)\Sigma_{n}+
\sum_{l=x,y,z}\mathcal{V}_l^{e}\left(\mathbf{r}\right)\Lambda_{l}\right],\nonumber\\
V^{o}\left(\mathbf{r}\right)=
\sum_{j=x,y}s_j\left[\sum_{n=x,y,z}\mathcal{U}_{jn}^{o}\left(\mathbf{r}\right)\Sigma_{n}+
\sum_{l=x,y,z}\mathcal{V}_{jl}^{o}\left(\mathbf{r}\right)\Lambda_{l}\right].
\label{eq:dis_s}
\end{gather}
Even terms conserve z-spin, whereas odd terms induce spin-flip. After projecting these terms onto a single band we obtain for inter-valley disorder potentials, Eq.~\eqref{eq:dis_v},
\begin{gather}
V_{c,v}^{int}\left(\mathbf{r}\right)=\mathbf{V}_{c,v}\left(\mathbf{r}\right)
\cdot\boldsymbol{\tau}, \mbox{ with}\nonumber\\
V_{c,v}^x\left(\mathbf{r}\right)=V_{yy}\mp V_{xx}-\frac{\gamma}{\Delta}\left(\partial_y V_{zx}\pm
\partial_x V_{zy}\right),\nonumber\\
V_{c,v}^y\left(\mathbf{r}\right)=-V_{yx}\mp V_{xy}-\frac{\gamma}{\Delta}\left(\partial_y V_{zy}\mp
\partial_x V_{zx}\right).
\label{eq:dis_1}
\end{gather}Similarly, for even and odd spin-dependent disorder potentials, Eqs.~\eqref{eq:dis_s}, we arrive at
\begin{widetext}
\begin{gather}
V_{c,v}^{e}\left(\mathbf{r}\right)=
\left(\mathcal{V}_{z}^e\pm \mathcal{U}_{z}^e\right)\tau_z s_z+
\frac{\gamma}{\Delta}s_z\left[\pm\left\{\mathbf{p},\boldsymbol{\mathcal{U}}^e\right\}
+\left[\nabla\times\boldsymbol{\mathcal{U}}^e\right]_z\tau_z
+\left(\pm\left\{p_y,\mathcal{V}_{y}^e\right\}-\left\{p_x,\mathcal{V}_{x}^e\right\}\right)\tau_x
-\left(\left\{p_x,\mathcal{V}_{y}^e\right\}\pm\left\{p_y,\mathcal{V}_{x}^e\right\}\right)\tau_y\right],\nonumber\\
 V_{c,v}^{o}\left(\mathbf{r}\right)=\sum_{j=x,y}s_j\left\{
\left(\mathcal{V}_{jz}^o\pm \mathcal{U}_{jz}^o\right)\tau_z+
\frac{\gamma}{\Delta}\left[\pm\left\{\mathbf{p},\boldsymbol{\mathcal{U}}_j^o\right\}
+\left[\nabla\times\boldsymbol{\mathcal{U}}_j^o\right]_z\tau_z
+\left(\pm\left\{p_y,\mathcal{V}_{jy}^o\right\}
-\left\{p_x,\mathcal{V}_{jx}^o\right\}\right)\tau_x-
\right.\right.\nonumber\\\left.\left.
-\left(\left\{p_x,\mathcal{V}_{jy}^o\right\}\pm\left\{p_y,\mathcal{V}_{jx}^o\right\}\right)\tau_y\right]\right\}.
\label{eq:dis_2}
\end{gather}
\end{widetext}
Equations~\eqref{eq:dis_1}-\eqref{eq:dis_2} enable us to relate parameters in the phenomenological model for disorder in Eq.~\eqref{eq:dis} to their microscopic counterparts as\begin{gather}
u_z\left(\mathbf{r}\right)=\mathcal{V}_z^e\pm\mathcal{U}_z^e+\frac{\gamma}{\Delta}\left[\nabla\times
\boldsymbol{\mathcal{U}}^2\right]_z,\nonumber\\
\mathbf{a}_{gz}=\frac{\gamma}{\Delta}\boldsymbol{\mathcal{U}}^e,\nonumber\\
u_{sf}^i=\mathcal{V}_{iz}^o\pm \mathcal{U}_{iz}^o+\frac{\gamma}{\Delta}\left[\nabla\times\boldsymbol{\mathcal{U}}_i^o\right]_z,\nonumber\\
\mathbf{w}_{\alpha}=\frac{\gamma}{\Delta}\boldsymbol{\mathcal{U}}_{\alpha}^o,\nonumber\\
\mathbf{u}_i=\mathbf{V}_{c,v},\nonumber\\
\mathbf{w}_{zx}=\frac{\gamma}{\Delta}\left(-\mathcal{V}_x^e,\pm\mathcal{V}_y^e\right),\nonumber\\
\mathbf{w}_{zy}=\frac{\gamma}{\Delta}\left(-\mathcal{V}_y^e,\mp\mathcal{V}_x^e\right),\nonumber\\
\mathbf{w}_{\alpha x}=\frac{\gamma}{\Delta}\left(-\mathcal{V}_{\alpha x}^o,\pm\mathcal{V}_{\alpha y}^o\right),\nonumber\\
\mathbf{w}_{\alpha y}=\frac{\gamma}{\Delta}\left(-\mathcal{V}_{\alpha y}^o,\mp\mathcal{V}_{\alpha x}^o\right).
\end{gather}


\begin{thebibliography}{60}
\bibitem{footnotemass}
We neglect a possible effective mass difference \cite{2DBands5} between two Kramers doublets ($K_+,\uparrow; K_-,\downarrow$) and ($K_+,\downarrow; K_-,\uparrow$).
\bibitem{foot}
For spinful electrons at valleys $K_{\pm}$ time inversion is implemented by the anti-unitary operator $\mathcal{T}=is_y\tau_x\mathcal{K}$, where $\mathcal{K}$ stands for complex conjugation. Therefore, $s_i \tau_j \xrightarrow{\;\mathcal{T}\;} -s_i\tau_j$ for $i,j=x,y$. As consequence, terms in the Hamiltonian that could simultaneously flip spin and scatter an electron in the same Wannier state between $K_+$ and $K_-$ are forbidden since their appearence would require breaking time-inversion symmetry, as it is deduced from Tab.~\ref{tab:electronic_operators_main}. More generally, time-inversion symmetry forbids spin-flip scattering of electrons in the same Wannier state between any pair of opposite points in the Brillouin zone. An illustration of this general statement, given in Ref.~\onlinecite{Yafet}, appeals to a free 2D electron scattering off a portential $V\left(\mathbf{r}\right)$, for SO coupling is described by $\mathcal{H}_{SO}\propto \left[\nabla V\left(\mathbf{r}\right)\times\mathbf{p}\right]\cdot\mathbf{s}$. For such a pertubation, spin-flip scattering has an amplitude $\left\langle\mathbf{k}\alpha|\mathcal{H}_{SO}|\mathbf{k}'\beta\right\rangle\propto i\left(\mathbf{k}\times\mathbf{k}'\right)\cdot\left[\mathbf{s}\right]_{\alpha\beta} V\left(\mathbf{k}-\mathbf{k}'\right)$, which vanishes for the pairs of states related by the time-inversion transformation ($V\left(\mathbf{q}\right)$ is the Fourier transform of $V\left(\mathbf{r}\right)$). Hence, $\left|\left\langle-\mathbf{k}-\alpha|\mathcal{H}_{SO}|\mathbf{k}\beta\right\rangle\right|^2 \equiv 0$.
\bibitem{foot2}
With this choice, time inversion is implemented by the anti-unitary operator $\mathcal{T}=i\sigma_y\mathcal{K}$.
\bibitem{foot_gap} Note that $\Delta$ refers to the band gap, not to the spin splitting of the bands, $\lambda$.
\bibitem{berry}
This is the Berry curvature deduced from the $\mathbf{k}\cdot\mathbf{p}$ Hamiltonian neglecting warping effects.
\end{thebibliography}
\end{document}